\documentclass[a4paper,fleqn]{cas-dc}
\usepackage[numbers,sort&compress]{natbib}

\def\tsc#1{\csdef{#1}{\textsc{\lowercase{#1}}\xspace}}
\tsc{WGM}
\tsc{QE}
\tsc{EP}
\tsc{PMS}
\tsc{BEC}
\tsc{DE}
\begin{document}
\let\WriteBookmarks\relax
\def\floatpagepagefraction{1}
\def\textpagefraction{.001}
\shorttitle{Intricate Magnetism and Topological Hall Effect Observed in Cr$_{0.83}$Te}
\shortauthors{Shubham et~al.}
\title [mode = title]{Intricate Magnetic Interactions and Topological Hall Effect Observed in Itinerant Room-temperature Layered Ferromagnet Cr$_{0.83}$Te}
\author{Shubham Purwar}
\credit{Conceptualization, Methodology, Validation, Formal analysis, Investigation, Writing original draft, Writing review \& editing}
\affiliation{organization={S. N. Bose National Centre for Basic Sciences},
    addressline={Salt Lake, JD Block, Sector III, Bidhannagar},
    city={Kolkata},
    postcode={700106},
    state={West Bengal},
    country={India}}
\author{Susmita Changdar}
\credit{Writing review \& editing}
\author{Susanta Ghosh}
\credit{Writing review \& editing}
\author{Tushar Kanti Bhowmik}
\credit{Methodology, Validation, Writing review \& editing}

\author {Setti Thirupathaiah}[orcid=0000-0003-1258-0981]
\cormark[1]
\ead{setti@bose.res.in}
\ead[URL]{www.qmat.in}
\credit{Conceptualization, Validation, Resources, Supervision, Project administration, Funding acquisition, Writing review \&
editing}

\begin{abstract}
We report the magnetic, electrical, and magnetotransport (Hall effect) properties of the hexagonal itinerant ferromagnet Cr$_{0.83}$Te. Further, a comprehensive study of the magneto-entropy scaling behavior has been done around the Curie temperature of $T_C \approx$ 338 K. A maximum entropy change (-$\Delta S_{m}^{max}$) of 2.77 $J/kg-K$ and relative cooling power (RCP) of 88.29 $J/kg$ near the T$_C$ have been achieved under an applied magnetic field of 5 Tesla. The critical exponents, $\beta$ = 0.4739(4), $\gamma$ = 1.2812(3), and $\delta$ = 3.7037(5), have been extracted using the magneto-entropy scaling analysis. The obtained critical exponents indicate the presence of intricate magnetic interactions in Cr$_{0.83}$Te. On the other hand, the magnetotransport study reveals a topological Hall effect attributed to the noncoplanar spin structure coexisting with a robust magnetocrystalline anisotropy. Further, we observe that the extrinsic skew-scattering mechanism originated anomalous Hall effect. Our experimental findings of the anomalous and topological Hall effect properties in the presence of intriguing high-temperature itinerant ferromagnetism and magnetocaloric effect in Cr$_{0.83}$Te can offer potential technological applications at room temperature.
 \end{abstract}


\maketitle

\section{Introduction}

Investigation of the low dimensional magnetic materials with ferromagnetic ordering at room temperature and strong magnetocrystalline anisotropy~\cite{liu2022interlayer,soumyanarayanan2016emergent} has gained a lot of research interests in recent days due to their potential applications in magnetic refrigeration and spintronic devices~\cite{deng2018gate,wang2019mnx,geim2013van, PhysRevLett.107.217202,khan2020recent,hossain2022synthesis}.  Hence, the van der Waals (vdW) ferromagnets are of great research interest from the fundamental science and advanced technology point of view due to their peculiar two-dimensional (2D) magnetic properties~\cite{gong2017discovery,seyler2018ligand,fei2018two,gibertini2019magnetic} and strong magnetocrystalline anisotropy~\cite{PhysRevB.93.134407}. Usually, the Heisenberg-type ferromagnet does not exist with intrinsic long-range magnetic ordering at finite temperature in the 2D limit due to dominant thermal fluctuations~\cite{PhysRevLett.17.1133}. Nevertheless, the single domain anisotropy or the exchange anisotropy can overcome the thermal fluctuation and allow the long-range magnetic ordering in 2D ferromagnets~\cite{Huang2017}. In this way, the 2D ferromagnetism has been found experimentally in many vdW materials such as Cr$_2$Ge$_2$Te$_6$ (T$_C\approx$ 61 K)~\cite{PhysRevB.96.054406}, Cr$_2$Si$_2$Te$_6$ (T$_C\approx$ 32 K)~\cite{PhysRevB.92.144404,xie2019two}, Fe$_3$GeTe$_2$ (T$_C\approx$ 215 K)~\cite{fei2018two,feng2022anisotropy}, and CrI$_3$ (T$_C \approx$ 45 K)~\cite{Huang2017},  but the long-range FM ordering temperature is far below the room temperature, limiting their usage in technological applications. Since a very few systems such as MnP (T$_C\approx$ 303 K) show room temperature 2D ferromagnetism in the bulk phase~\cite{Sun2020}, searching for new room-temperature layered FM materials coupled with large magnetocrystalline anisotropy is crucial for realizing potential technological applications.

Furthermore, the FM vdW materials posses another peculiar property such as the magnetocaloric effect (MCE). The  magnetocaloric effect (MCE) at room temperature in the FM vdW materials with maximum entropy change -$\Delta S_{m}^{max}$  is of recent research interest due to their technological applications in the environmental friendly magnetic refrigerators~\cite{magnetochemistry7050060}. There exists quite a few 2D FM systems showing significant -$\Delta S_{M}^{max}$ such as Fe$_{3-x}$GeTe$_2$ (1.1 $J/kg-K$ at 5 T)~\cite{doi:10.1021/acs.inorgchem.5b01260},  Cr$_5$Te$_8$ (1.6 $J/kg-K$ at 5 T)~\cite{PhysRevB.100.245114}, Cr$_2$Si$_2$Te$_6$ (5.05 $J/kg-K$ at 5 T), and  Cr$_2$Ge$_2$Te$_6$ (2.64 $J/kg-K$ at 5 T)~\cite{PhysRevMaterials.3.014001}.

Theoretical studies suggest that the layered Cr$_x$Te$_y$ systems are the potential candidates to realize the much-anticipated room-temperature 2D ferromagnetism in bulk~\cite{PhysRevMaterials.2.081001}. Since then, a variety of Cr$_x$Te$_y$ compounds have been grown experimentally and studied for their peculiar 2D ferromagnetism, including CrTe~\cite{ETO200116}, Cr$_2$Te$_3$~\cite{wang2018ferromagnetic}, Cr$_3$Te$_4$~\cite{hessen1993hexakis}, Cr$_4$Te$_5$~\cite{Zhang2020,liu2023magnetic,wang2022fabrication},  and Cr$_5$Te$_8$~\cite{PhysRevB.100.024434, luo2018magnetic}. Generally, Cr$_x$Te$_y$ compounds possess alternating stacks of CrTe$_2$ layers intercalated by the Cr layers (excess) along the $\it{z}$-axis~\cite{dijkstra1989band}. The intercalated Cr concentration plays an important role in the formation of these compounds' crystal structure, magnetic structure, and transport properties~\cite{Huang2004, Huang2006,Wontcheu2008, IPSER1983265}. The topological Hall effect observed in Cr$_x$Te$_y$ systems implies the presence of non-trivial spin texture such as skyrmions and bi-skyrmions~\cite{zhang2023room,li2022air}  or the noncoplanar magnetic structures~\cite{chen2023observation,Purwar2023,huang2021possible}.

On the other hand, though there exist a couple of studies on the hexagonal Cr$_{1-x}$Te type systems, discussing the magnetic, transport, and Hall effect properties~\cite{liu2022magnetic, he2020large}, no systematic study is available on the magnetocrystalline anisotropy (MCA) and magneto-entropy scaling analysis which offer a better understanding on the topological Hall effect and magnetic exchange interactions, respectively. Importantly, the magnetic exchange interactions are profoundly influenced by the MCA, manifesting the Heisenberg, XY, Ising, or complex magnets~\cite{kaul1985static, BedoyaPinto2021, Lee2021a}. In addition, the strong MCA is crucial for engineering the non-coplanar spin-structure that leads to generating the skyrmion lattice~\cite{Preissinger2021, Low2022}. On the other hand, the magneto-entropy scaling analysis deduces the critical exponents, defining the strength and the type of magnetic interactions present in the system~\cite{PhysRevB.100.245114, PhysRevMaterials.3.014001, PhysRevB.96.134410, Liu2023}.

   \begin{figure*}[t!]
    \centering
    \includegraphics[width=\linewidth]{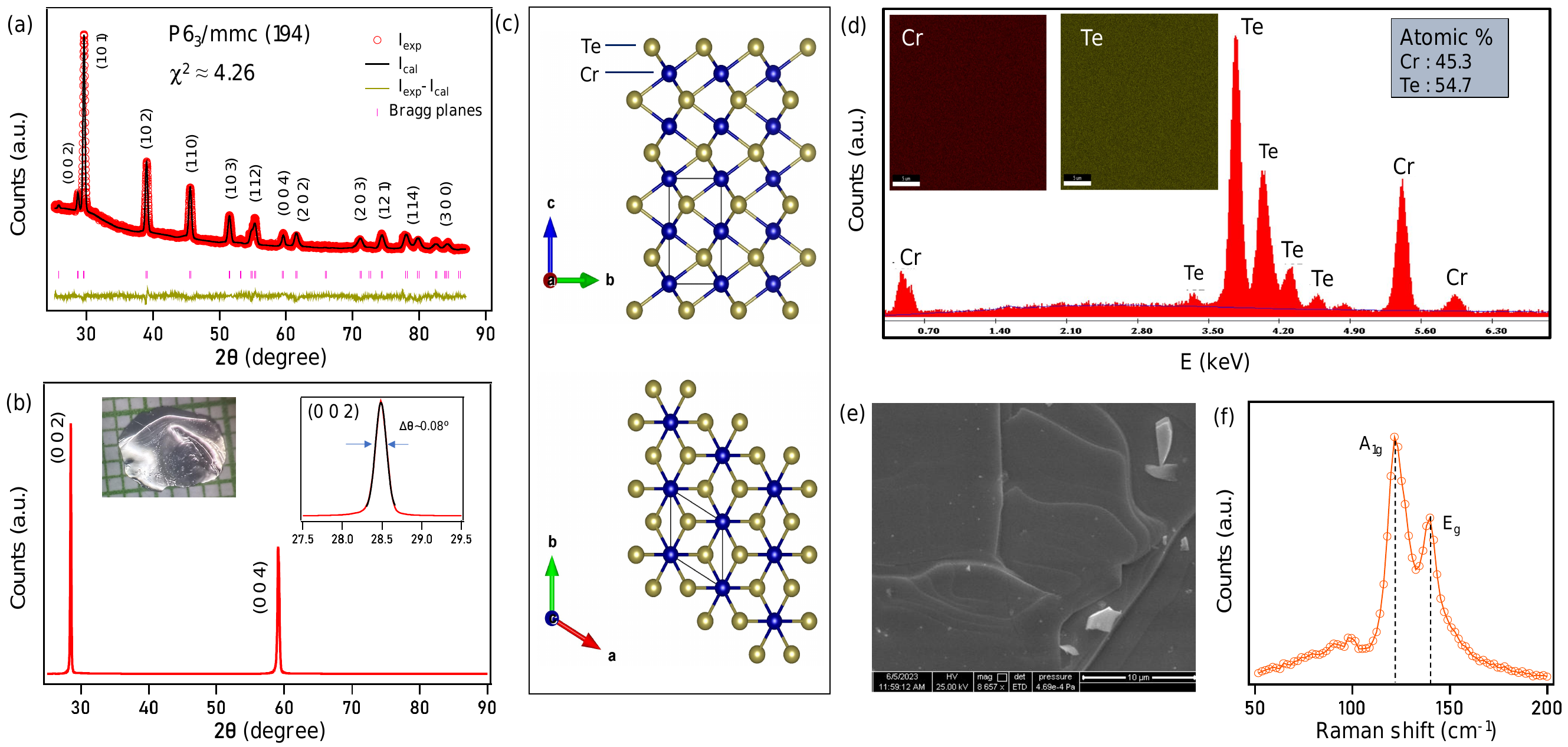}
    \caption{(a) Powder x-ray diffraction (XRD) of the crushed Cr$_{0.83}$Te single crystals overlapped with Rietveld refinement. (b) XRD pattern of Cr$_{0.83}$Te single crystal. The left inset of (b) shows the photographic image of a typical Cr$_{0.83}$Te single crystal. The right inset of (b) shows the rocking curve of (0 0 2) Bragg's reflection with FWHM of 0.08$^{\circ}$. (c) Schematics of the CrTe hexagonal crystal structure projected onto the $bc$-plane and $ab$-plane. (d) Energy dispersive X-ray spectroscopy (EDXS) spectra of Cr$_{0.83}$Te single crystal. Top-insets of (d) show the elemental mapping of Cr and Te. (e) Scanning electron microscopy (SEM) image of Cr$_{0.83}$Te single crystal. (f) Raman spectra of Cr$_{0.83}$Te single crystal.}
    \label{1}
\end{figure*}

Therefore, this study reports the anisotropic magnetic properties, anomalous and topological Hall effects, and magnetocaloric effect in the hexagonal Cr$_{0.83}$Te single crystal. Cr$_{0.83}$Te is an interesting system as it falls in between the van der Waals trigonal (vdW) CrTe$_2$~\cite{Zheng2023} and non-vdW hexagonal CrTe~\cite{Wu2021}, despite all three being layered systems. As the intercalated Cr atoms play a vital role in shaping the magnetic and magnetotransport properties, Cr$_{0.83}$Te could be a potential candidate to show the topological Hall effect (THE) originating from the noncoplanar spin structure of the intercalated Cr spins. Moreover, it is one of the few Cr$_x$Te$_y$ type systems showing room temperature ferromagnetism, idle for room temperature technological applications such as magnetic storage devices~\cite{Sbiaa2011},  spin valves~\cite{Cortie2020},  magnetic tunnel junctions~\cite{Cortie2020}, and spin-transfer torque devices~\cite{Krizakova2021}. Although an earlier study on the similar composition of hexagonal Cr$_{0.833}$Te (Cr$_5$Te$_6$) performed critical analysis but lacked details on the relation between the anisotropic magnetic properties and the topological Hall effect~\cite{zhang2022multiple}. Another study performed magnetotransport measurements on a different crystal structure of monoclinic thin film Cr$_{0.833}$Te, yet lacks the discussion on the relation between MCA and THE~\cite{chen2023observation}. Therefore, this study aims to unravel the relation between the magnetocrystalline anisotropy,  the topological Hall effect, and the magnetic exchange interactions in the hexagonal Cr$_{0.83}$Te. Further, the re-scaled magnetoentropy change, -$\Delta S_m (T, H)$, exhibits a remarkable convergence onto a universal curve, suggesting a second-order magnetic transition in this systems~\cite{oesterreicher1984magnetic,law2018quantitative}. Extracted critical exponents from the field-dependent magnetoentropy change highlight the interplay between the 3D-Ising and the meanfield-type exchange interactions.

\section{Experimental details}
High-quality single crystals of Cr$_{0.83}$Te were grown using the chemical vapor transport (CVT) method with iodine as a transport agent. We thoroughly mixed  Cr (99.99\%, Alfa Aesar) and Te (99.99\%, Alfa Aesar) powders in a $5:5$ ratio, and a small quantity of iodine (3 mg/cc) was added to the powder mixture. The mixture was sealed in a quartz tube under argon gas and placed in a gradient two-zone horizontal tube furnace for about 15 days. One end of the tube was heated at 1000$^{\circ}$C (source) and the other end was kept at 820$^{\circ}$C (sink), following a previously established procedure~\cite{hashimoto1971magnetic}. The obtained single crystals were large, measuring up to  $5\times5$ mm$^2$, and looking shiny. A representative photographic image of a typical single crystal is shown in the inset of Fig.~\ref{1}(b).

The crystal structure of as-grown Cr$_{0.83}$Te single crystals was examined by the x-ray diffraction (XRD) technique using the Rigaku x-ray diffractometer (SmartLab, 9 kW) with Cu K$_\alpha$ radiation of wavelength of $\lambda=$1.5406 \AA. We employed scanning electron microscopy (SEM) and energy-dispersive X-ray spectroscopy (EDXS) techniques to explore surface morphology and elemental compositions. Magnetic and transport properties were studied by the physical property measurement system (9 Tesla-PPMS, DynaCool, Quantum Design). Electrical resistivity and Hall measurements were conducted using the conventional four-probe technique. To eliminate the influence of longitudinal magnetoresistance, caused mainly by the voltage probe misalignment, the Hall resistivity was determined as $\rho_{xy}(H) = [\rho_{xy}(+H)-\rho_{xy}(-H)]/2$. In addition, Raman spectra were captured using a micro-Raman spectrometer (LabRam HR Evolution HORIBA France SAS) equipped with a 532 nm laser.

\begin{figure}[t]
    \centering
    \includegraphics[width=\linewidth]{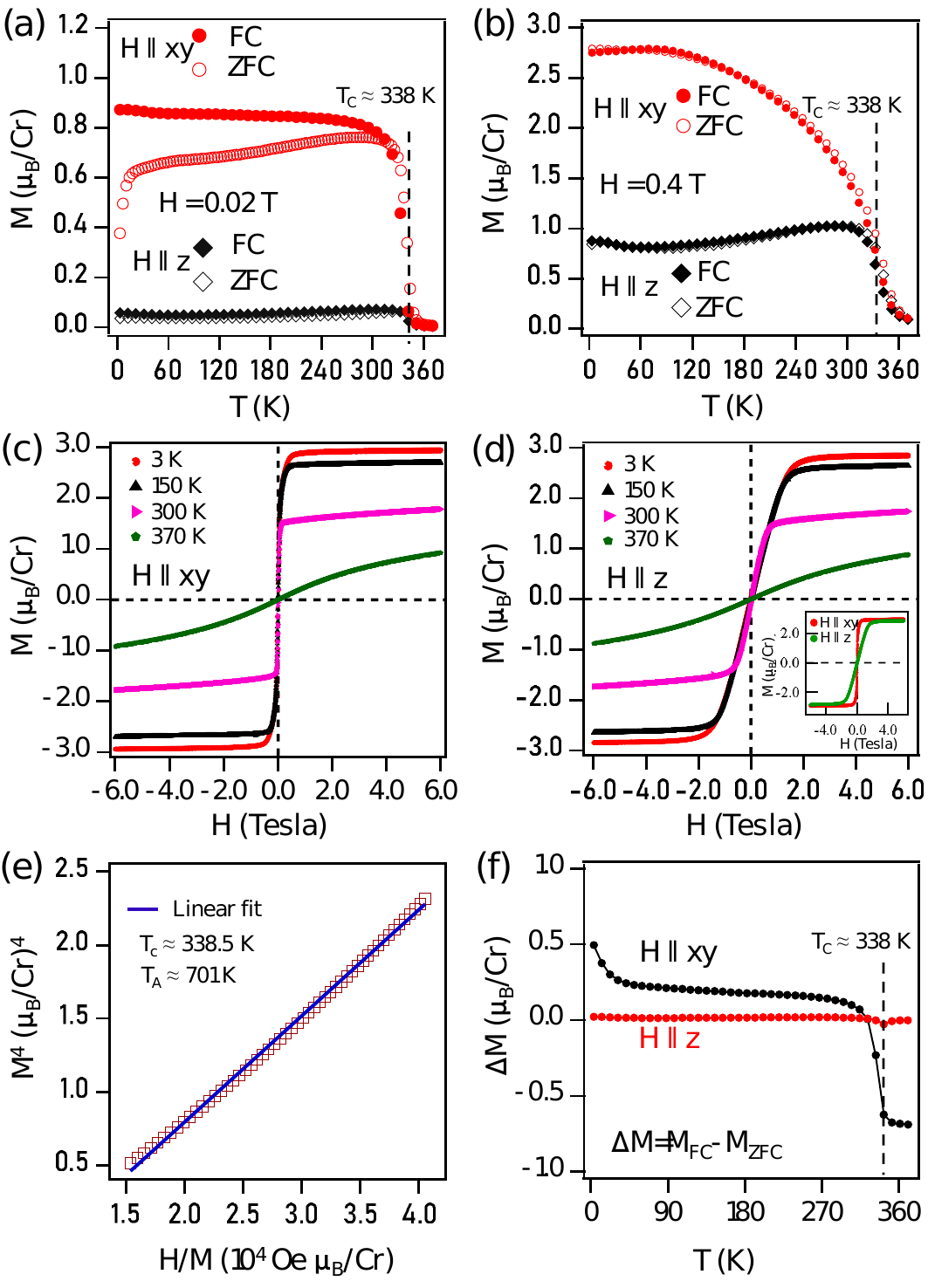}
    \caption{Temperature-dependent magnetization [$M(T)$] of Cr$_{0.83}$Te taken in zero-field-cooled (ZFC) and field-cooled (FC) modes with magnetic fields H = 0.02 T (a) and 0.4 T (b) for $H\parallel \it{xy}$ and $H\parallel \it{z}$.  Magnetization isotherms [$M(H)$] measured at various sample temperatures for (c) $H\parallel \it{xy}$ and (d) $H\parallel \it{z}$. (e) Plot of $M^4$ $\it{vs}$ $H/M$. (f)  Temperature-dependent $\Delta M$ ($M_{FC} - M_{ZFC}$) derived for $H\parallel xy$ and $H\parallel z$.}
    \label{2}
\end{figure}

\section{Results and Discussion}

\subsection{Structural Properties}
Figure~\ref{1}(a) displays the XRD pattern of crushed Cr$_{0.83}$Te single crystals taken at room temperature, confirming the hexagonal crystal structure with P6$_3$/mmc space group. No additional impurity phases were detected. Rietveld refinement, overlapped on the XRD data, yields lattice parameters $a = b = 3.9808 (4)$ \AA ~and $c = 6.2122(3)$ \AA, which are very close to the previous reports on similar systems~\cite{ido1969exchange,liu2022magnetic}. Fig.~\ref{1}(b) depicts the XRD pattern of Cr$_{0.83}$Te single crystal, showing the intensity of $(0~0~\it{l})$ Bragg plane, suggesting that the crystal growth axis is along the $c$-axis. The rocking curve of (0 0 2) plane shown in the right-inset of Fig.~\ref{1}(b) displays a single sharp peak with a full width at half maximum (FWHM) of $\Delta\theta$ = 0.08$^\circ$, confirming the high quality of single crystals~\cite{PhysRevB.107.144425}. The crystal structure of Cr$_{0.83}$Te, as schematically presented in the top-panel of Fig.~\ref{1}(c), reveals stacking of CrTe$_2$ layers along the $c$-axis arrangement without significant vdWs gap between two CrTe$_2$ layers. This arrangement differs from other Cr$_x$Te$_y$-based systems in which a significant vdW gap is present~\cite{Purwar2023, Zheng2023}. From the crystal structure projected onto the $ab$-plane, as shown in the bottom-inset of  Fig.~\ref{1}(c), we can observe an intertwined honeycomb lattice consisting of Cr and Te atoms. Fig.~\ref{1}(d) depicts the EDXS spectra, revealing the Cr:Te atomic ratio of 0.83:1, consistent with the chemical composition of Cr$_{0.83}$Te obtained from the powder XRD refinement. The EDXS mappings shown in the insets of Fig.~\ref{1}(d) confirm the uniform distribution of Cr and Te elements in the studied single crystal. The obtained sample composition of Cr$_{0.83}$Te hints at the 17\% of Cr vacancies. These vacancies seem critical in maintaining the NiAs-type hexagonal crystal structure~\cite{dijkstra1989band,liu2022magnetic,li2022diverse}. Fig.~\ref{1}(e) presents the SEM image of Cr$_{0.83}$Te single crystal with terraces of the different layers, demonstrating the layered nature of the system. The Raman spectra of Cr$_{0.83}$Te as shown in Fig.~\ref{1}(f) unveil two prominent phonon peaks, positioned at approximately 123.5 cm$^{-1}$ and 139.79 cm$^{-1}$. These peaks correspond to the distinct vibrational modes, the out-of-plane A$_{1g}$ and the in-plane E$_g$, of the Cr$_x$Te$_y$ system~\cite{chen2022air,li2022diverse}.

\subsection{Magnetic Properties}
\subsubsection{Magnetization Measurements}
To explore the magnetic properties of Cr$_{0.83}$Te, we performed magnetization measurements as a function of temperature [$M(T)$] and field [$M(H)$]. The magnetic field was applied parallel to both the $\it{xy}$-plane [$H\parallel \it{xy}$] and the $\it{z}$-axis [$H\parallel \it{z}$]. Figs.~\ref{2}(a) and \ref{2}(b) exhibit the in-plane ($H\parallel \it{xy}$) and out-of-plane ($H\parallel \it{z}$) $M(T)$ curves obtained under magnetic fields of 0.02 T and 0.4 T, respectively. As we can notice, the system undergoes a paramagnetic-to-ferromagnetic (PM-FM) transition at a Curie temperature ($T_C$) of approximately 338 K, very close to the previously reported value on a similar system~\cite{liu2022magnetic}. The bifurcation in the $M(T)$ curves at $T_C$ between ZFC and FC, as observed in Fig.~\ref{2}(a), can be ascribed to the magnetic ordering, which is thermally irreversible at lower fields (0.02 T) due to canted magnetic moments~\cite{zhang2020tunable, HUANG20081099}. This interpretation is supported by the $M(T)$ data taken at 0.4 T of the applied field [see Fig.~\ref{2}(b)], in which we can see the absence of bifurcation between ZFC and FC data at the $T_C$ due to complete alignment of the canted moments along the field direction. Further, at an applied field of 0.02 T,  a sharp downturn in the ZFC $M(T)$ data is noticed at around 15 K, which disappears at 0.4 T, suggesting a possible spin-glass type transition below 15 K~\cite{Roy2015} at lower fields. Nevertheless, we notice significant magnetization anisotropy between in-plane and out-of-plane orientations at both applied fields.

Figs.~\ref{2}(c) and \ref{2}(d) depict the magnetization isotherms, $M(H)$, measured at different temperatures for $H\parallel \it{xy}$ and $H\parallel \it{z}$ orientations, respectively. The $M(H)$ data suggest Cr$_{0.83}$Te to be a soft-ferromagnet with negligible coercivity. The magnetic anisotropy observed from the $M(T)$ data is further verified by the magnetization isotherms by plotting in-plane and out-of-plane $M(H)$ at 3 K, as shown in the inset of Fig.~\ref{2}(d). The magnetic anisotropy observed in these systems plausibly stems from the non-coplanar magnetic structure resulting from the Cr vacancies~\cite{chen2023observation, Purwar2023}. From Figs.~\ref{2}(c) and \ref{2}(d), it is clear that Cr$_{0.83}$Te has an easy-axis parallel to the $\it{xy}$-plane. This observation is substantiated by the magnetization saturation occurring at an applied field of 0.5 T for $H\parallel \it{xy}$. In comparison, 1.9 T is needed for the same with $H\parallel \it{z}$ when measured at 3 K. Note here that an easy-axis of magnetization parallel to the $\it{z}$-axis was found from a similar system of Cr$_{0.87}$Te which has 4.6\% of higher Cr compared to our studied Cr$_{0.83}$Te, again confirming that the magnetic structure is highly sensitive to the Cr concentration present in these systems~\cite{Fujisawa2020,zhang2023room}. The saturation magnetization ($M_{s}$) for both in-plane and out-of-plane orientations is determined as 2.95 $\mu_B$/Cr and 2.86 $\mu_B$/Cr, respectively. This is notably smaller than the calculated ordered moment of Cr (3.4 $\mu_B$/Cr) from the band structure calculations due to the itinerant nature of the Cr-$d$ electrons~\cite{PhysRevB.53.7673,dijkstra1989band}.

Next, to estimate the degree of itinerant ferromagnetism in Cr$_{0.83}$Te,  we employed Takahashi's self-consistent renormalization (SCR) theory around T$_C$~\cite{Takahashi1986}. According to the SCR theory, the magnetization $M$ and the magnetic field $H$ at $T_C$ are related by,

\begin{equation}
  M^4 = \frac{1}{4.671} \left[\frac{T_C^2}{T_A^3}\right] \left(\frac{H}{M}\right)
 \label{Eq1}
 \end{equation}

Here, $T_A$ denotes the dispersion of the spin fluctuation spectrum in the wave-vector space. Fig.~\ref{2}(e) shows the M$^4$ $\it{vs.}$ $H/M$ for $H \parallel \it{xy}$, fitted nicely by the linear Eq.~\ref{Eq1}. This linear relationship is generally observed in itinerant ferromagnetic systems like LaCo$_2$P$_2$ ($T_C/T_0\approx0.14$)~\cite{PhysRevB.91.184414}, Fe$_4$GeTe$_2$ ($T_C/T_0\approx0.16$)~\cite{Mondal2021}, SmCoAsO ($T_C/T_0\approx0.12$)~\cite{PhysRevB.82.054421}, and Cr$_4$Te$_5$ ($T_C/T_0\approx0.063$)~\cite{liu2023magnetic}. The fit yields a slope of 7.2168$\times$10$^{-5}$ [$\mu_B$/Cr]$^5$/Oe. Using the slope and T$_C$ values, we estimate T$_A$ to be approximately 701 K for $H\parallel \it{xy}$. As per the SCR theory, the $T_C$ can be described by,

\begin{equation}
  T_C = (60c)^{-3/4}M_{sp}^{3/2}T_A^{3/4}T_0^{1/4}
 \label{Eq2}
 \end{equation}

Here, $c=$0.3353, $M_{sp}$ represents the spontaneous magnetization, and T$_0$ denotes the energy width of the dynamical spin fluctuation spectrum. Using the values of $T_C$, $M_{sp}$, and $T_A$, we deduce the characteristic temperature $T_0$ = 4963 K for Cr$_{0.83}$Te. Further, the SCR spin fluctuation theory suggests that the ratio T$_C$/T$_0$ defines the degree of itineracy in the ferromagnets. Such as, the spin moments are localized for T$_C$/T$_0\approx$1 and delocalized for $T_C/T_0\ll$ 1~\cite{takahashi2013spin}. In Cr$_{0.83}$Te single crystals, we estimate T$_C$/T$_0$ $\approx$ 0.07 ($\ll$ 1),  confirming the itinerant ferromagnetic behavior.

Fig.~\ref{2}(f) depicts $\Delta M$($M_{FC}$-$M_{ZFC}$) plotted as a function of temperature for both $H\parallel \it{z}$ and $H\parallel \it{xy}$. From Fig.~\ref{2}(f) it is evident that the out-of-plane magnetization does not change much between ZFC and FC modes, while the $\Delta M$ of in-plane magnetization is very sensitive at around $T_C$. More importantly,  $\Delta M$ rapidly increases with decreasing temperature below 40 K.  This kind of magnetic behavior could stem from multiple factors, including the inherent magnetic anisotropy of these systems~\cite{liu2022magnetic,PhysRevB.100.024434} or the noncoplanar magnetic structure resulting from the Cr vacancies~\cite{chen2023observation,Purwar2023}.

\subsubsection{Magnetocrystalline Anisotropy}

\begin{figure}[t]
    \centering
   \includegraphics[width=\linewidth]{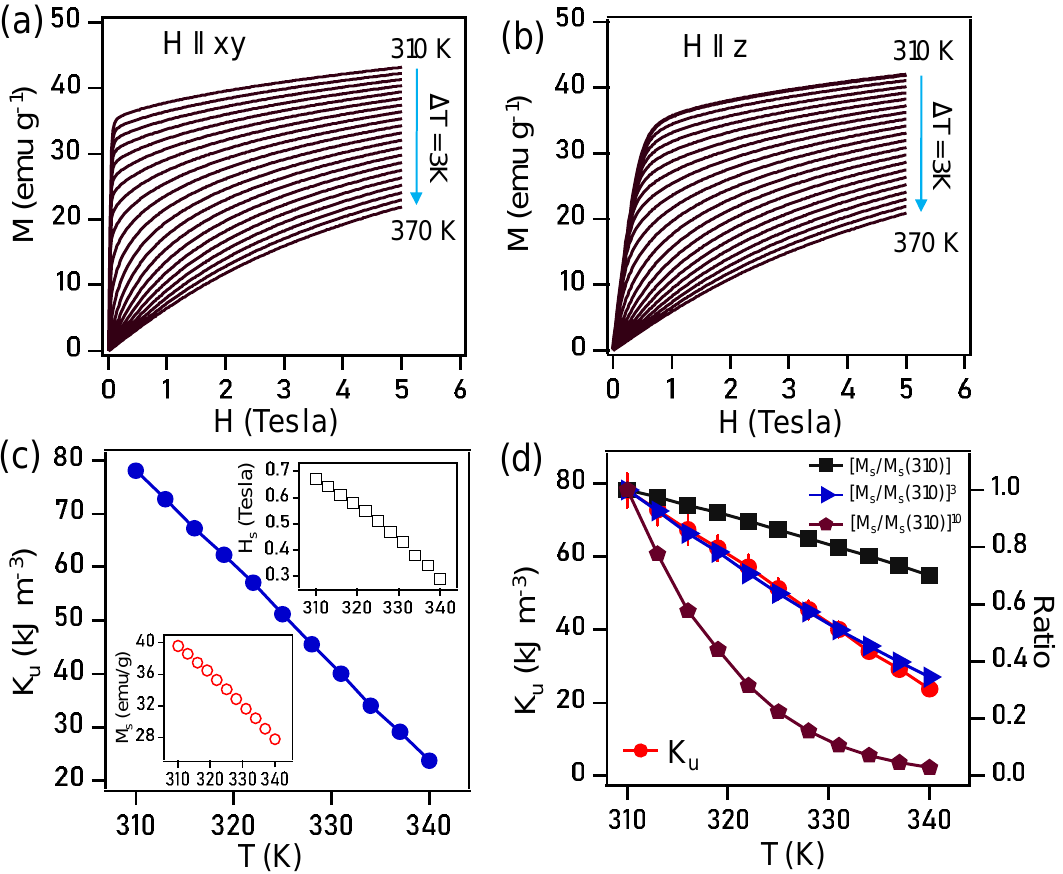}
   \caption{Magnetization isotherms measured around $T_C$ for (a) $H\parallel \it{xy}$ and (b) $H\parallel \it{z}$. (c) Temperature-dependent in-plane magnetocrystalline anisotropy energy density $K_u$. The bottom inset in (c) presents saturation magnetization ($M_s$) and the top inset shows saturation magnetic field ($H_s$) estimated below $T_C$. (d) Ratios of [M$_s$/M$_s$(310 K)]$^{n(n+1)/2}$ (right-axis) overlaid with magnetocrystalline anisotropy (left-axis) for $n$ = 1, 2, and 4.}
   \label{5}
\end{figure}

With the help of magnetization isotherms [$M(H)$] as shown in the Figs.~\ref{5}(a) and \ref{5}(b), we studied the magnetocrystalline anisotropy (MCA) energy density ($K_u$) using the expression~\cite{li2022air},
\begin{equation}
 K_u = \mu_0\int_{0}^{M_s} [H_{xy}(M)-H_{z}(M)]dM
 \label{Eq0}
 \end{equation}
 Here $M_s$ denotes the saturation magnetization. $H_{\it{xy}}$ and $H_{\it{z}}$ represent the fields applied along $\it{xy}$ and $\it{z}$ directions, respectively. Fig.~\ref{5}(c) depicts the temperature-dependent $K_u$ derived from the experimental data. We find $K_u$ = 78.11 kJ/m$^3$ at T = 310 K, gradually decreasing with increasing temperature and reaching 23.73 kJ/m$^3$ at $T_C$ (338 K). On the other hand, the estimated K$_u$ $\approx$ 390 kJ/m$^3$ at 40 K [see Fig.~\ref{4}(d)] is much larger than the K$_u$ values reported on many 2D magnetic systems such as CrBr$_3$~\cite{PhysRevMaterials.2.024004}, Cr$_2$Ge$_2$Te$_6$~\cite{PhysRevMaterials.3.014001}, and Cr$_2$Si$_2$Te$_6$~\cite{PhysRevMaterials.3.014001} and comparable to the K$_u$ values of CrI$_3$~\cite{PhysRevMaterials.2.024004} and Fe$_4$GeTe$_2$~\cite{Seo2020, Mondal2021}. Though Fe$_3$GeTe$_2$ shows a large magnetocrystalline anisotropy of 1460 kJ/m$^3$ at 2 K~\cite{leon2016magnetic}, its Curie temperature is much below the room temperature (T$_C$ = 220 K)~\cite{leon2016magnetic}. Thus, Cr$_{0.83}$Te having a large K$_u$ value of $\approx$ 390 kJ/m$^3$ with a Curie temperature of 338 K seems to be a promising candidate from the technological applications point of view.  See Table~\ref{T1} for a list of 2D materials and their respective magnetocrystalline anisotropy energies ($K_u$).

\begin{figure*}[hbt!]
    \centering
    \includegraphics[width=0.95\linewidth]{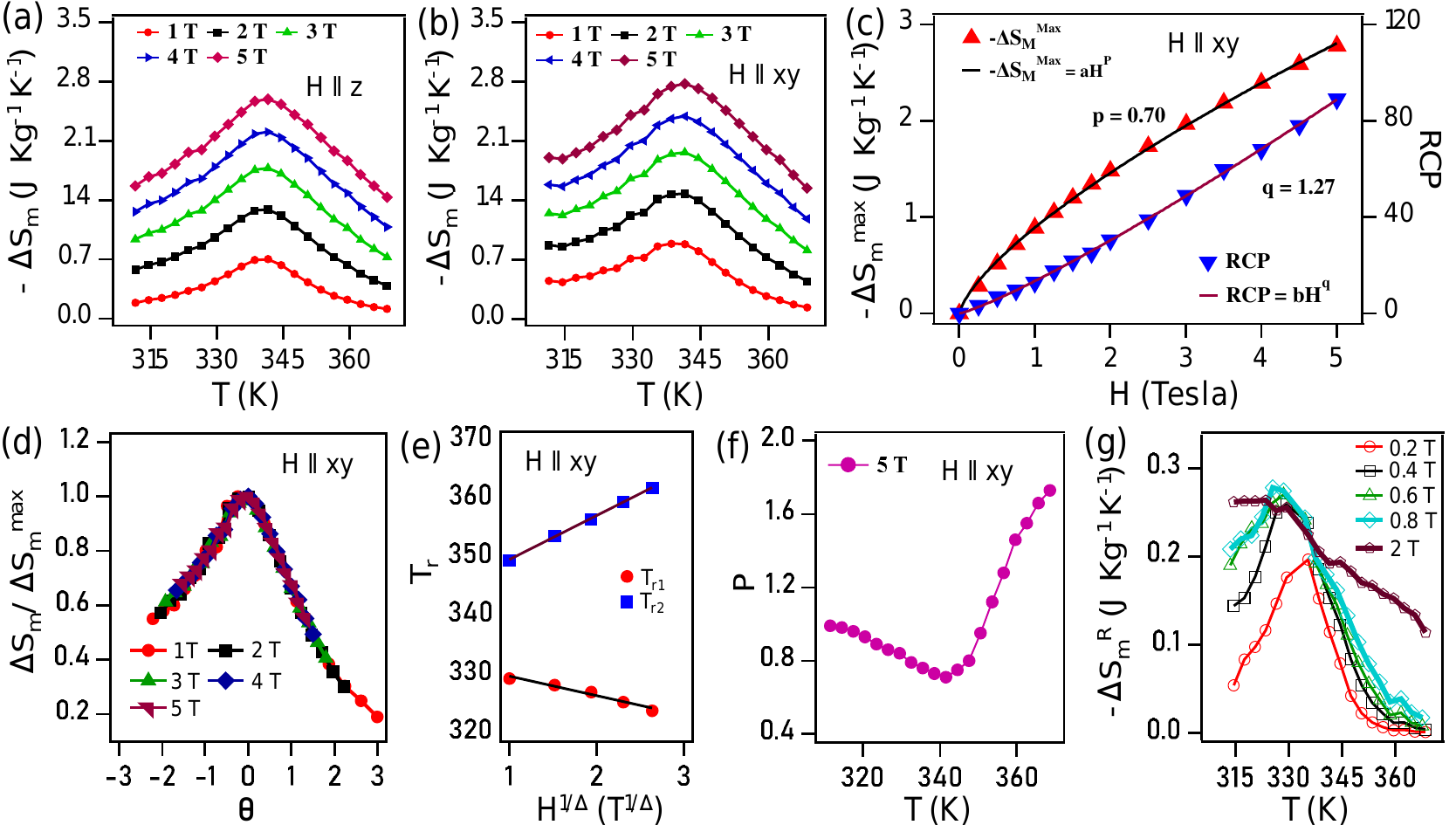}
    \caption{Magnetic entropy change -$\Delta S_m$ plotted as a function of temperature at different magnetic fields for (a) $H\parallel \it{z}$ and (b) $H\parallel \it{xy}$. (c) Field-dependent maximum magnetic entropy change (-$\Delta S_m^{max}$) (left axis) and relative cooling power (RCP) (right axis). (d) Normalized magnetic entropy change as a function of rescaled temperature $\theta$ for various applied fields. (e) $T_r$ $\it{vs}$ $H^{1/\Delta}$. (f) Exponent $p$ plotted as a function of temperature. (g) Rotational magnetic entropy change (-$\Delta S_m^R$) plotted as a function of temperature.}
    \label{6}
\end{figure*}

\begin{table}[h]
\caption{Magnetocrystalline anisotropy energies (K$_u$) of different layered materials.}
\begin{tabular*}{\linewidth}{c @{\extracolsep{\fill}} cccc}
 \hline
  Composition & $\approx$ K$_u$ (in kJ/m$^3$) & Ref. \\ [1.5ex]
 \hline\hline
 Cr$_{0.83}$Te & 390 &  This work \\ [1.0ex]
\hline
 Cr$_{0.87}$Te & 270 &  \cite{liu2022magnetic} \\ [1.0ex]
\hline
Cr$_{0.69}$Te & 165 & \cite{Purwar2023} \\[1.0ex]
 \hline
Cr$_{0.625}$Te & 94 & \cite{PhysRevB.100.245114} \\[1.0ex]
 \hline
 Cr$_{0.6}$Te & 174 & \cite{huang2021possible} \\[1.0ex]
 \hline
 CrBr$_3$ & 86 & \cite{PhysRevMaterials.2.024004}\\ [1.0ex]
 \hline
  CrI$_3$ & 300 &  \cite{PhysRevMaterials.2.024004}\\[1.0ex]
 \hline
Cr$_2$Ge$_2$Te$_6$ & 20 & \cite{PhysRevMaterials.3.014001}\\ [1.0ex]
 \hline
 Cr$_2$Si$_2$Te$_6$ & 65 &  \cite{PhysRevMaterials.3.014001}\\[1.0ex]
 \hline
 Fe$_4$GeTe$_2$ & 250 &  \cite{Seo2020}\\[1.0ex]
 \hline
  Fe$_3$GeTe$_2$ & 1460 &  \cite{leon2016magnetic}\\[1.0ex]
 \hline
  \hline
\end{tabular*}
\label{T1}
\end{table}

Since the magnetic anisotropy expectation value $<K^n>$ is directly proportional to $M_s^{n(n+1)/2}$, as per the classical theory of magnetism~\cite{PhysRev.96.1335,carr1958temperature}, we plotted $[\frac{M_{s}(T)}{M_{s}(310)}]^{n(n+1)/2}$ for n = 1, 2, and 4. Here, n = 1 represents intrinsic anisotropy, n = 2 represents uniaxial anisotropy, and n = 4 represents cubic anisotropy, giving rise to the exponents 1, 3, and 10, respectively. In Fig.~\ref{5}(d), we plotted $[M_{s}(T)/M_{s}(310)]$,  $[M_{s}(T)/M_{s}(310)]^3$, and $[M_{s}(T)/M_{s}(310)]^{10}$ ratios as a function of temperature. Most importantly, in Fig.~\ref{5}(d), the overlapped K$_u$(T) of Fig.~\ref{5}(c) matches very well with $[M_{s}(T)/M_{s}(310)]^3$ ratio, confirming the dominant uniaxial anisotropy in this system that is strongly temperature dependent. The temperature-dependent K$_u$ originates from the fluctuating local spin clusters activated from the thermal energy~\cite{PhysRev.96.1335,carr1958temperature}.

\subsubsection{Magnetocaloric Effect}

To investigate the magnetocaloric effect (MCE), we analyzed the field-dependent isotherms $M(H)$ acquired at different temperatures [see Figs.~\ref{5}(a) and \ref{5}(b)] in the vicinity of $T_C$ for both $H\parallel \it{xy}$ and $H\parallel \it{z}$ orientations. The magnetocaloric effect is an intrinsic property of a ferromagnetic system, resulting from adiabatic heating or cooling under external magnetic fields~\cite{PECHARSKY199944}. This effect leads to the generation of magnetic entropy change $\Delta S_m (T, H)$, which can be quantified using the formula,

\begin{equation}
\Delta S_m (T, H) =  \int_{o}^{H} (\frac{\partial S}{\partial H})_T dH = \int_{o}^{H} (\frac{\partial M}{\partial T})_H dH
 \label{Eq9}
 \end{equation}

where ($\frac{\partial S}{\partial H})_T$ = ($\frac{\partial M}{\partial T})_H$  based on Maxwell’s relation. For the magnetization data acquired at smaller discrete field and temperature intervals, the magnetic entropy change $\Delta S_m (T, H)$ can be expressed as

\begin{equation}
\Delta S_m (T, H) = \frac{\int_{0}^{H} M(T_{i+1}, H) dH - \int_{0}^{H} M(T_{i}, H) dH }{T_{i+1}-T_i}
 \label{Eq10}
\end{equation}

\begin{table*}
\caption{Critical exponents of Cr$_{0.83}$Te single crystal compared with several theoretical models (MEC = Magnetic Entropy Change).}
\begin{tabular*}{\linewidth}{c @{\extracolsep{\fill}} cccccccc}
 \hline\hline
 Composition  & Technique & $\beta$ & $\gamma$ & $\delta$ & p & q & Ref. \\ [1.5ex]
 \hline\hline
Cr$_{0.83}$Te  & MEC & 0.4739(4) & 1.2812(3) & 3.7037(5) & 0.70 & 1.27 & This work \\ [1.2ex]

\hline
Landau mean field  &Theory & 0.5 & 1 & 3 & 0.667 & 1.333 & \cite{PhysRev.108.1394} \\[1.2ex]
 \hline
 3D Heisenberg  & Theory & 0.365 & 1.386 & 4.80 & 0.637 & 1.208 & \cite{kaul1985static}\\ [1.2ex]
 \hline
 3D Ising  & Theory & 0.325 & 1.241 & 4.82 & 0.569 & 1.207 & \cite{kaul1985static}\\[1.2ex]
 \hline\hline
\end{tabular*}
\label{T2}
\end{table*}

Figs.~\ref{6}(a) and \ref{6}(b) show - $\Delta S_m (T, H)$ plotted as a function temperature under various magnetic fields up to 5 T taken with a step size of 1 T for both $H\parallel \it{xy}$ and $H\parallel \it{z}$ orientations, respectively. All - $\Delta S_m (T, H)$ curves show a maximum change in entropy with a broad peak at around $T_C$ as can be seen from Figs.~\ref{6}(a) and \ref{6}(b). Further, we observe that the value of - $\Delta S_m^{max} (T, H)$ increase monotonically with field for $H\parallel \it{xy}$ [see Fig.~\ref{6}(c)]. Under an applied field of 5 T, the maximum of -$\Delta S_m (T, H)$ is about 2.78 J kg$^{-1}$ K$^{-1}$ for $H\parallel \it{xy}$ and is about 2.58 J kg$^{-1}$ K$^{-1}$  for $H\parallel \it{z}$. These -$\Delta S_m (T, H)$ values taken at 5 T are comparable to the other 2D ferromagnetic systems such as Cr$_2$Ge$_2$Te$_6$ (2.64 J kg$^{-1}$ K$^{-1}$)~\cite{PhysRevMaterials.3.014001} and Cr$_5$Te$_8$ (2.38 J kg$^{-1}$ K$^{-1}$)~\cite{PhysRevB.100.245114}, larger than the values of Fe$_{3-x}$GeTe$_2$ (1.14 J kg$^{-1}$ K$^{-1}$)~\cite{Liu2019} and CrI$_3$ (1.56J kg$^{-1}$ K$^{-1}$)~\cite{Liu2018}, and smaller than the values of CrB$_3$ (7.2 J kg$^{-1}$ K$^{-1}$)~\cite{Yu2019} and Cr$_2$Si$_2$Te$_6$ (5.05 J kg$^{-1}$ K$^{-1}$)~\cite{PhysRevMaterials.3.014001}.


To estimate the relative cooling power (RCP) as shown in Fig.~\ref{6}(c), we employed the relation RCP = -$\Delta S_m^{max}$ $\times$ $\delta T_{FWHM}$, where -$\Delta S_m^{max}$ is the maximum entropy change near $T_C$ and $\delta T_{FWHM}$ is the full width at half maximum of the peak~\cite{gschneidner1999recent}. The calculated RCP in Cr$_{0.83}$Te is 88.29 J kg$^{-1}$ at around $T_C$ with an applied field of 5 T parallel to the $\it{xy}$-plane. The RCP value of Cr$_{0.83}$Te obtained in this study is comparable to the RCP value obtained in the other 2D systems such as Cr$_2$Ge$_2$Te$_6$ (87 J kg$^{-1}$)\cite{PhysRevMaterials.3.014001}, but smaller than the values obtained from Cr$_5$Te$_8$ (131.2 J kg$^{-1}$)~\cite{PhysRevB.100.245114}, CrI$_3$ (122.6 J kg$^{-1}$)~\cite{Liu2018}, Cr$_2$Si$_2$Te$_6$ (114 J kg$^{-1}$)~\cite{PhysRevMaterials.3.014001}, Fe$_{3-x}$GeTe$_2$(113 J kg$^{-1}$)~\cite{Liu2019}, and CrBr$_3$(191.5 J kg$^{-1}$)~\cite{Yu2019}.

In addition, both -$\Delta S_m^{max}$ and RCP are related by the power law of magnetic field as given below~\cite{gschneidner1999recent,franco2006field},
\begin{equation}
  -\Delta S_m^{max} = aH^p
   \label{Eq11}
 \end{equation}
 \begin{equation}
      RCP =  bH^q
       \label{Eq12}
 \end{equation}
 where p and q are the exponents. At T = T$_C$,  they can be written as
 \begin{equation}
     p = 1 + \frac{\beta - 1}{\beta + \gamma}
      \label{Eq13}
 \end{equation}
 \begin{equation}
     q = 1 + \frac{1}{\delta}
      \label{Eq14}
 \end{equation}

Where $\beta$, $\gamma$, and $\delta$ are critical exponents, which can be found using alternative theoretical models such as isothermal analysis, these exponents obtained from the magnetic entropy analysis, however, are more dependable without the use of initial exponents. In addition, the exponent $\delta$ has been calculated using the Widom scaling relation, $\delta$ = 1 + $(\gamma$/$\beta)$~\cite{widom1964degree}. The fit of -$\Delta S_m^{max}$ by the Eq.~\ref{Eq11} yields $p$ = 0.70. Similarly, the field dependence of RCP is fitted by Eq.~\ref{Eq12}, yielding $q$ = 1.27. Based on Eq.~\ref{Eq13} and  Eq.~\ref{Eq14}, the derived critical exponents are $\beta$ = 0.4739(4), $\gamma$ = 1.2812(3), and $\delta$ = 3.7037(5). Note that no single conventional universality class can describe the derived critical exponents. The critical exponents suggest a crossover between the mean-field model ($\beta$ = 0.5) and the 3D-Ising model ($\gamma$ = 1.241). Thus, our study indicates complex magnetic interactions in the Cr$_{0.83}$Te system.

Next, we performed the scaling analysis of MCE following the procedure given by Franco \textit{et. al.} ~\cite{franco2006field,franco2007constant}. The scaling analysis of - $\Delta S_m (T, H)$ is constructed by normalizing the - $\Delta S_m (T, H)$ curves with respect to the maximum of -$\Delta S_m^{max}$ [$\frac{\Delta S_m (T, H)}{\Delta S_m^{max}}$]. The reduced temperature ($\theta_{\mp}$) is defined by choosing two reference temperatures ($T_{r1} \le T_C$ and $T_{r2}> T_C$), satisfying the condition, $\frac{\Delta S_m (T_{r1} < T_c)}{\Delta S_m^{max}}$  = $\frac{\Delta S_m (T_{r2} > T_c)}{\Delta S_m^{max}}$ = h. Here, h is a scaling constant with values within the $0 < h < 1$ range. Then, the rescaled temperature $\theta_{\mp}$ can be written as

 \begin{equation}
     \theta_{-}  = (T_C - T)/(T_{r1} - T_C),T \le T_C
      \label{Eq15}
 \end{equation}
 \begin{equation}
    \theta_{+} = (T - T_C)/(T_{r2} - T_C), T > T_C
      \label{Eq15}
 \end{equation}

\begin{figure*}
    \centering
    \includegraphics[width=\linewidth]{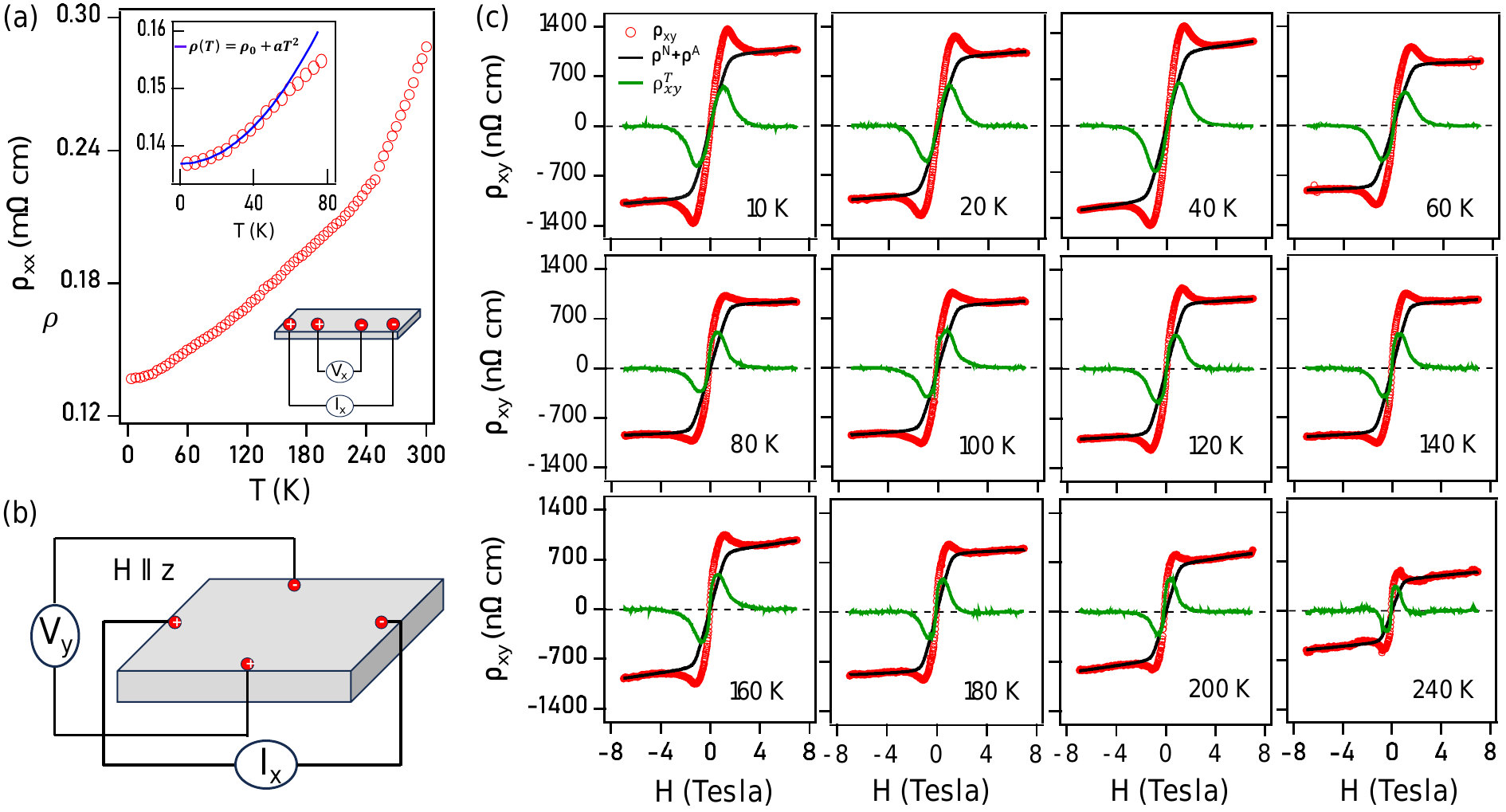}
    \caption{(a) Temperature-dependent longitudinal electrical resistivity $\rho_{xx}$. The top inset of (a) displays low-temperature resistivity fitted by $\rho (T) = \rho_{0} + aT^2$ and the bottom inset of (a) shows a schematic of linear four-probe geometry. (b) Schematic diagram of the Hall measurement geometry. (c) Transverse resistivity $\rho_{xy}$ (Hall resistivity), $\rho_{xy}$ plotted as a function of the magnetic field measured at various temperatures. In (c), the red curves represent total Hall resistivity (raw data), the black curves are the fittings using the equation $\rho_{N}+\rho_{A}$, and the green curves represent the topological Hall resistivity ($\rho_{T}$). See the text for more details.}
    \label{3}
\end{figure*}

Following the MCE scaling analysis, all the curves of $\frac{\Delta S_m}{\Delta S_m^{max}}$ plotted as a function of reduced temperature $\theta$ at various magnetic fields collapse into a single curve as shown in Fig.~\ref{6}(d), confirming the second-order magnetic phase transition in Cr$_{0.83}$Te~\cite{oesterreicher1984magnetic,law2018quantitative,franco2017predicting}. Moreover, it can be seen that the universal curve of MCE is independent of the applied magnetic field and temperature, as it is generally determined by the intrinsic magnetization of the system~\cite{franco2006field}. Further, the reference temperatures $T_{r1}$ and $T_{r2}$ linearly depend on $H^{1/\Delta}$ for $\Delta$ = $\beta$ + $\gamma$ as shown in Fig.~\ref{6}(e). Fig.~\ref{6}(f) displays the temperature dependence of $p$. The $p(T)$ curve follows universal behavior across $T_C$ as it reaches the value 1 for T $<$ $T_C$~\cite{FRANCO20091115}. On the other hand, well above $T_C$, $p$ reaches the value of two due to the Curie-Weiss law~\cite{FRANCO20091115}. At T = $T_C$, $p(T)$ has a minimum value of 0.7, in line with the universal law of $p(T)$~\cite{oesterreicher1984magnetic}. Overall, the temperature dependence of $p$ perfectly follows the universal behavior of a second-order phase transition~\cite{franco2006field}. The rotational magnetic entropy changes ($\Delta S_m^R$) can be calculated using the formula $\Delta S_m^R(T, H)$ = $\Delta S_m(T, H_{ab})$ - $\Delta S_m(T, H_{c})$. Fig.~\ref{6}(g) depicts the temperature dependence of - $\Delta S_m^R(T, H)$. From Fig.~\ref{6}(g), we can find a maximum of -$\Delta S_m^R(T, H)$ at around T$_C$ when derived at a field interval of 0.2 T. The maxima of -$\Delta S_m^R(T, H)$ shifts to lower temperatures with increasing field intervals, and for 2 T of field interval, no maxima is found down to 315 K. This behavior is generally found in  systems with strong magnetic anisotropy~\cite{PhysRevB.100.245114}.

\subsection{Electrical and Magnetotransport Properties}

Figure~\ref{3}(a) displays temperature-dependent longitudinal resistivity $\rho_{\it{xx}}$ of Cr$_{0.83}$Te single crystal, revealing the metallic nature throughout the measured temperature range~\cite{PhysRevB.98.195122}. A change in the resistivity curve is noticed at around 245 K, while the exact origin is not very clear to us, but we noticed a small cusp in the $dM/dT$ (not shown) at around 245 K from the FC data measured with 0.02 T [see Fig.~\ref{2}(a)]. The top inset of Fig.~\ref{3}(a) elucidates the quadratic temperature dependence of $\rho_{\it{xx}}$,  fitted by $\rho(T)=\rho_0 + aT^2$, indicating dominant electron-electron scattering at low temperatures. This observation is in agreement with the theoretical studies on weak itinerant ferromagnetic metals in which a T$^2$ dependence of $\rho_{\it{xx}}$ was predicted at low temperatures~\cite{ueda1975}. The bottom inset of Fig.~\ref{3}(a) depicts the linear four-probe geometry used for measuring the longitudinal resistivity. The transverse resistivity $\rho_{\it{xy}}$ (Hall resistivity) as a function of the applied field ($H$) is shown in Fig.~\ref{3}(c) measured at various sample temperatures. The Hall resistivity, $\rho_{\it{xy}}$, was measured with the current applied along the $\it{x}$-axis, the magnetic field applied along the $\it{z}$-axis, and the resulting  Hall voltage was detected along the $\it{y}$-axis as demonstrated in Fig.~\ref{3}(b). Thus, the red-colored curves shown in the panels of Fig.~\ref{3}(c) exhibit the raw data of field-dependent Hall resistivity $\rho_{xy}$ recorded at various sample temperatures, and the black-colored curves are the fittings to the total Hall resistivity using the formula \cite{Nagaosa2010},

\begin{equation}
    \rho_{\it{xy}}(H) = \rho^N(H)+\rho^A(H) = \mu_0R_0H+\mu_0R_sM
    \label{eq1}
\end{equation}

Here, $\rho^N$ and $\rho^A$ are the normal Hall and anomalous Hall contributions to the total Hall resistivity, respectively. $R_0$ is the normal Hall coefficient, $R_s$ is the anomalous Hall coefficient, and $M$ is the isothermal magnetization. The normal and anomalous Hall coefficients can be determined by a linear fit at the high field region following the relation $\frac{\rho_{\it{xy}}}{\mu_0H}=R_0+\frac{R_sM}{H}$, as shown in Fig.~S\ref{1} of the supplemental information. However, as can be seen from the panels of Fig.~\ref{3}(c), the fitting of $\rho_{\it{xy}}$ with Eq.~\ref{eq1} is not perfect due to the topological Hall contribution. Thus, by including the topological Hall effect (THE) contribution, the total Hall resistivity can be expressed by $\rho_{\it{xy}}(H) = \mu_0R_0H+\mu_0R_sM + \rho^{T}$ and the topological Hall resistivity ($\rho^{T}$) is extracted using the relation $ \rho^{T} = \rho_{\it{xy}}(H)-(\mu_0R_0H+\mu_0R_sM)$ ~\cite{PhysRevLett.106.156603, Purwar2023}. In Fig.~\ref{3}(c), the green-colored curves represent the topological Hall resistivity.

\begin{figure}[t]
    \centering
    \includegraphics[width=\linewidth]{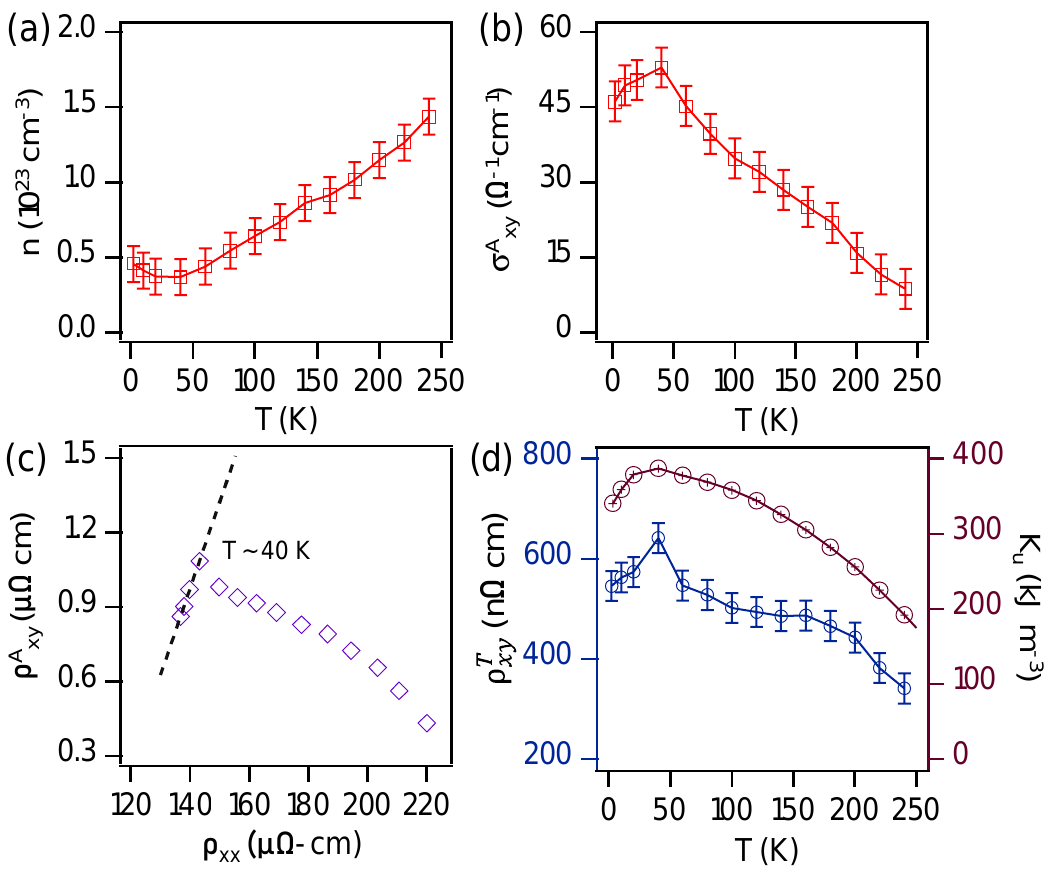}
    \caption{Temperature dependence of (a) derived calculated carrier density n. (b) Anomalous Hall conductivity $\sigma^A_{\it{xy}}$. (c) Plot of $\rho^A_{\it{xy}}$ vs. $\rho_{\it{xx}}$ with linear fitting up to $40 K$. (d) Maximum value of topological Hall resistivity $\rho_{\it{xy}}^{T, max}$ (left axis) and magneto-crystalline anisotropy density, $K_u$ (right axis), plotted as a function of temperature.}
    \label{4}
\end{figure}

Figure~\ref{4}(a) depicts the charge-carrier density ($\it{n}$) plotted as a function of temperature, estimated from the normal Hall coefficient $R_0$ following the relation $\it{n}=1/(R_0q$). Here $q$ is the hole-carrier charge. The derived carrier concentration $\it{n}$ decreases with increasing temperature up to 40 K. However, above 40 K, it increases monotonously with temperature. In Cr$_x$Te$_y$ systems, the competition between the FM and AFM phases plays a vital role in the magnetotransport properties~\cite{liu2022magnetic,chen2023observation, Purwar2023}. Similarly, in Fig.~\ref{2}(f), the in-plane $\Delta M(T)$ changes abruptly around T$\approx$ 40 K, plausibly originating the non-monotonic changes of carrier concentration around 40 K. Next, the anomalous Hall conductivity (AHC), $\sigma_{\it{xy}}$, obtained using the relation $\sigma_{\it{xy}}=\frac{\rho_{\it{xy}}}{\rho_{\it{xy}}^2 +\rho_{\it{xx}}^2 }$ is presented in Fig.~\ref{4}(b). $\sigma_{\it{xy}}$ increases with increasing temperature up to 40 K. Beyond 40 K, it monotonically decreases with increasing temperature. A maximum $\sigma_{\it{xy}}$ $\approx$ 52 $\Omega^{-1}cm^{-1}$ is found at around 40 K. In general, the anomalous Hall conductivity can intrinsically arise from the electronic structure-originated non-zero Berry curvature~\cite{Nagaosa2010, Zeng2006} or extrinsically due to magnetism-originated side-jump/skew-scattering mechanisms~\cite{Smit1958, Nagaosa2010}.

To understand the nature of AHE in Cr$_{0.83}$Te we plotted $\rho_{\it{xy}}$ $\it{vs.}$ $\rho_{\it{xx}}$ in Fig.~\ref{4}(c). The Hall resistivity for an itinerant ferromagnetic metal is generally expressed by the formula $\rho_{\it{xy}} = \alpha\rho_{\it{\it{xx}}} + \beta\rho_{\it{xx}}^2$, where $\alpha$ and $\beta$ represent the skew-scattering and side-jump coefficients, respectively~\cite{Nagaosa2010, PhysRevB.2.4559}. From Fig.~\ref{4}(c), a linear dependence of $\rho_{\it{xy}}$ on $\rho_{\it{xx}}$ is evident up to 40 K, beyond which a deviation occurs with increasing temperature. This observation hints at the skew-scattering mechanism that originated anomalous Hall effect in Cr$_{0.83}$Te, consistent with the other Cr$_x$Te$_y$ systems~\cite{Purwar2023, PhysRevB.98.195122, PhysRevB.102.144433}. Furthermore, Fig.~\ref{4}(d) depicts the maximum amplitude of topological Hall resistivity ($\rho_{\it{xy}}^{T, max}$) (left axis) plotted as a function of temperature. $\rho_{\it{xy}}^{T, max}$ increases with temperature and reaches a maximum value of approximately 620 n$\Omega$-cm at around 40 K, consistent with $\Delta M(T)$ ($H\parallel \it{xy}$) of Fig.~\ref{2}(f). The right axis of Fig.~\ref{4}(d) displays the magnetocrystalline anisotropy energy constant, $K_u$, calculated using Eq.~\ref{Eq0}. It is worth mentioning here that the critical point of 40 K observed from $\rho^{A}_{xy}$, $\rho^{T}_{xy}$, and $K_u$ (see Fig.~\ref{4}) originates from a canted antiferromagnetic transition noticed at around 40 K (from $M(T)$ measured with 2 T, not shown), which is consistent with the other Cr$_x$Te$_y$ based systems~\cite{li2022diverse, HUANG20081099}.

Several theories have been proposed to elucidate the origin of the chiral spin texture that can manifest the topological Hall effect. These theories include the Dzyaloshinskii-Moriya interaction (DMI) in noncentrosymmetric systems under a strong spin-orbit coupling~\cite{Yi2009, PhysRevLett.110.117202, wang2022topological, shao2019topological, PhysRevLett.106.156603}, geometrically frustrated magnetic interactions~\cite{Machida2007}, and noncoplanar spin structure stabilized by the magnetocrystalline anisotropy (MCA)~\cite{Low2022, Purwar2023}. On the other hand, the intricate magnetism observed in Cr$_{x}$Te$_y$ type systems results from a complex interplay of magnetic interactions, including direct, super-exchange, and double exchange mechanisms, noncollinear magnetism, and a mixed valence state of the Cr ions~\cite{chen2023observation,zhang2022multiple,zhang2019high}. The derived critical exponents, which deviate from any single universality class (as discussed above), strongly suggest the presence of robust magnetocrystalline anisotropy (MCA) in Cr$_{0.83}$Te. Consequently, the large MCA can stabilize the requisite chiral spin structure~\cite{Wang2019, Preissinger2021}. Thus, in the presence of chiral-spin structure, the itinerant electrons acquire real-space Berry curvature, leading to a non-zero scalar-spin chirality $\chi_{ijk} = S_{i}(S_{j}\times S_k)\neq 0$ to produce the topological Hall effect~\cite{Machida2007,wang2022topological}. The analogy of chiral-spin structure stabilized by the large MCA is consistent with our experimental data shown in Fig.~\ref{4}(d) (right-axis), where the temperature-dependent topological Hall resistivity ($\rho_{\it{xy}}^T$) replicates the temperature dependence of MCA energy density $K_u$, having the maximum $\rho_{\it{xy}}^T$ at the maximum K$_u\approx390$ kJ/m$^3$ at 40 K.

\section{Summary}
We investigated the intricate magnetism, electrical, and magnetotransport properties in the hexagonal itinerant ferromagnet Cr$_{0.83}$Te. The magnetotransport study reveals a substantial topological Hall effect, originating from the noncoplanar spin structure in the presence of strong magnetic anisotropy and the skew-scattering-induced anomalous Hall effect. We observe a remarkable cooling efficacy of -$\Delta S_{m}^{max}$ $\approx$ 2.77 J $kg^{-1}$$K^{-1}$ and RCP $\approx$ 88.29 J/kg at an applied field of 5 Tesla. By utilizing the magneto-entropy scaling analysis, we extracted the critical exponents $\beta$ = 0.4739(4), $\gamma$ = 1.2812(3), and $\delta$ = 3.7037(5), which do not follow any single universality class, suggesting a complex magnetic interaction in Cr$_{0.83}$Te. Re-scaled -$\Delta S_m (T, H)$ curves fall into a single universal curve, confirming the second-order magnetic transition. The spin-fluctuation parameter, derived from the critical magnetization isotherms based on the SCR theory, confirms the itinerant ferromagnetic nature of Cr$_{0.83}$Te.

\section{Acknowledgements}
S.C. and S.G. acknowledge the University Grants Commission (UGC), India, for the PhD fellowship. Part of this work has been done using instruments from the Technical Research Centre (TRC) of the S. N. Bose National Centre for Basic Sciences, established under the TRC project of the Department of Science and Technology (DST), Govt. of India.

\bibliographystyle{model1-num-names}
\bibliography{CrTe}

\end{document}